\documentclass[dvipdft]{IEEEtran}
\pdfoutput=1

\IEEEoverridecommandlockouts
\usepackage{cite}
\usepackage{amsmath,amssymb,amsfonts}
\usepackage{algorithmic}
\usepackage{textcomp}
\usepackage{xcolor}
\usepackage{soul}
\usepackage{csquotes}
\usepackage{ulem}
\usepackage{verbatim}

\usepackage{tabularx}
\usepackage{subcaption}

\usepackage{pdfpages}

\usepackage{bbm,dsfont}

\usepackage{verbatim}
\def\BibTeX{{\rm B\kern-.05em{\sc i\kern-.025em b}\kern-.08em
\kern-.1667em\lower.7ex\hbox{E}\kern-.125emX}}
    
\usepackage{xcolor}
\definecolor{verde}{rgb}{0.25,0.5,0.35}
\definecolor{jpurple}{rgb}{0.5,0,0.35}
\definecolor{darkgreen}{rgb}{0.0, 0.2, 0.13}
\makeatletter
\def\ScaleIfNeeded{%
\ifdim\Gin@nat@width>\linewidth
\linewidth
\else
\Gin@nat@width
\fi
}
\makeatother
\begin{document}

\title{
A System and Methodology for Third-Party Evaluation of Bias in Sensitive Personal Information Used to Train Financial Models
}

\title{
%Evaluation of 
Analysis Bias in Sensitive Personal Information Used to Train Financial Models
}

\author{\IEEEauthorblockN{Reginald Bryant, Celia Cintas, Isaac Wambugu, Andrew Kinai, Komminist Weldemariam} \\
\IEEEauthorblockA{IBM Research | Africa, Nairobi, Kenya}
}
\begin{comment}

\author{\IEEEauthorblockN{1\textsuperscript{st} Given Name Surname}
\IEEEauthorblockA{\textit{dept. name of organization (of Aff.)} \\
\textit{name of organization (of Aff.)}\\
City, Country \\
email address}
\and
\IEEEauthorblockN{2\textsuperscript{nd} Given Name Surname}
\IEEEauthorblockA{\textit{dept. name of organization (of Aff.)} \\
\textit{name of organization (of Aff.)}\\
City, Country \\
email address}
\and
\IEEEauthorblockN{3\textsuperscript{rd} Given Name Surname}
\IEEEauthorblockA{\textit{dept. name of organization (of Aff.)} \\
\textit{name of organization (of Aff.)}\\
City, Country \\
email address}
%\and
%\and
%\and
}
\end{comment}

\IEEEoverridecommandlockouts
\IEEEpubid{\makebox[\columnwidth]{Preprint IEEEE Global Conference on Signal and Information Processing \hfill} \hspace{\columnsep}\makebox[\columnwidth]{ }}
\maketitle

\begin{abstract}
%TODO: Need to re-write the abstract.
Bias in data can have unintended consequences which propagate to the design, development, and deployment of machine learning models. In the financial services sector, this can result in discrimination from certain financial instruments and services. At the same time, data privacy is of paramount importance, and recent data breaches have seen reputational damage for large institutions. Presented in this paper is a trusted model-lifecycle management platform that attempts to ensure consumer data protection, anonymization, and fairness. Specifically, we examine how datasets can be reproduced using deep learning techniques to effectively retain important statistical features in datasets whilst simultaneously protecting data privacy and enabling safe and secure sharing of sensitive personal information beyond the current state-of-practice. 

%We conclude the paper by presenting a preliminary experimental evaluation of the proposed system, and its extension for the future. 
%The machine learning community is well aware of how data quality effects classifier performance. Specifically skewed training datasets are sure to create biased classifier predictions from unseen data--the so-called imbalanced learning problem. This problem is further exacerbated as many classifiers are trained on data which cannot be assessed by an impartial third-party for one reason or another. The aim of this research presented is to examine how datasets can be reproduced using Deep Learning (DL) techniques in order to effectively uncover levels of skewness in datasets whilst simultaneously protecting data privacy. Ultimately, these experiments are anticipated to lead to a trusted marketplace of models that can be scored based on sensitive training datasets which can be securely examined and assessed by third-parties.

\end{abstract}

\begin{IEEEkeywords}
bias, trust, machine learning models, financial services, data skewness
\end{IEEEkeywords}

\section{Introduction}

In spite of the potential advantages that the new digital transformations (e.g., the introduction of digital currencies, advances in AI) offer to several sectors ranging from financial services to healthcare, their advancement is still very much paralleled with risks. Chief amongst these pressing concerns is the secure and fair use of (user) data to provide diagnostics and access to services (e.g., lending, loan, credit, etc.). This -- if mismanaged -- could leave back doors open for both intentional and unintentional biases that can be exploited for unlawful acts. 
Fairness can happen at any level of data, model, algorithm and application stack, including within the underlying platform %, data, and framework 
\cite{%abs-1810-01943,Roselli:2019:MBA:3308560.3317590,
aif360-2018}.

Unlawful acts which are propagated, for example, in financial models in order to optimize return on investment, is mitigated due to customer protection. This is due to perceived financial service practices created by these financial models. There are also compliance requirements which are used to manage operational practices at the corresponding banks.
%due to perceived financial services practices created by financial models. 
Hence, financial institutions who wish to manage their reputations to avoid brand or reputational damage closely monitor the operational practices of and compliance of their correspondence banks as well as newly-acquired banks. 
For example, 
%TODO: Use another example
%This need was 
recently a high profile case was highlighted against the First National Bank (FNB) of South Africa  \cite{FNB}. The Usury Act was levied against FNB after acquiring the smaller Saambou Bank ---a bank that had operational difficulties managing their R8 billion (\$550 million USD) worth of mortgages.  

Thus, it becomes imperative across many domains and sectors that a robust, trusted platform to ensure 
end-to-end consumer data protection, anonymisation, and fairness exists. This is the main objective of 
this paper. In particular, in this paper, we describe our proposed distributed trusted model platform for 
small-to-medium business blockchain-based business networks.  We then discuss our novel methodology for sharing SPI (sensitive personal information) data beyond the traditionally used anonymization and data sharing techniques. 
Finally, we present our preliminary experimental evaluation of the proposed data synthesis techniques, 
demonstrating the utility (e.g., validation, verification, etc.) of our methodology and its usage in protecting 
consumer data. Finally, we present the analysis and summary of this work.

%In this paper, to make headway and begin to contribute to the body of work to further anonymization techniques and develop robust, trusted platforms for consumer data protection, we make the following contributions: 
%\begin{itemize}
    %\item developed a blockchain-based distributed trusted model platform for SME business networks;
    %\item Create a novel methodology to share SPI data (sensitive personal information) that furthers existing annonymization and sharing techniques; 
    %\item developed  novel methodology to share SPI data (sensitive personal information) beyond traditionally used annonymization and data sharing techniques; 
    %\item Provide experimental evaluation of the platform, and data synthesis and show our methodology works without scarifying privacy or accuracy;
    %\item experimentally evaluated the proposed platform and data synthesis techniques without scarifying privacy or accuracy;
   % \item And finally, we demonstrate the utility (e.g., validation, verification, etc.) of our methodology and its usage in protecting consumer data.
%    \item finally, demonstrated the utility (e.g., validation, verification, etc.) of our methodology and its usage in protecting consumer data.
%\end{itemize}

%In the remainder of the paper, we describe the underlying motivation for this work, outline the architecture and implementation of our trusted model platform, present the experiment and evaluation set-up and results, and finally present the analysis and summary of this work and its implications for future work.

\section{Motivation}
\label{sec.motivation}

%{\color{red}{TODO - In order of the listing:}
%\begin{itemize}
 %   \item Abdi: to review this section and ensure it is related to FSS stories.
  %  \item Regi/Andrew/Celia: to connect where the issues mentioned in this section addressed by TME and in our analysis … we need to relate/cross-reference between the sections.  
%\end{itemize}}

Several factors have motivated this work. First and foremost, this work is initially driven by the necessity to leverage distinctly small datasets to build machine learning (ML) models. This is the case in certain domains where there is a scarcity of data (low-frequency transactions).
These conditions require the expansion of  existing datasets to the level required by some ML algorithms, which we applied to use cases in financial services (e.g., credit scoring and credit-limit management \cite{SpeakmanSM18}). Data generation based on small datasets can become a powerful tool in the data scientist’s toolbox when considering these circumstances, and when working on network-based ML algorithms like Federated Learning where bespoke Deep Learning (DL) model structures need to be defined and optimized beforehand and later optimized as more real-world data is collected.

Secondly, as we later realized, the same same techniques selected for data expansion could be used for data protection. There is a need to develop novel ways to ensure data privacy that addresses the pitfall of existing techniques. Most financial institutions use popular state-of-practice anonymization techniques (e.g. removal, redaction, encryption, and data masking) to share data with other privileged institutions be it partners, vendors, or regulators. Unfortunately, these mechanisms are at high risk from bad actors as anonymization remains susceptible to de-identification \cite{rubinstein2016anonymization}. However, we find that with most DL synthesis techniques, no one-to-one relationship is formed between the real and synthesized datasets. This then makes decryption challenging to a degree which can be set prior to data generation.

% A mechanism for controlled data sharing that scores data utility and privacy.
Thirdly, trust involves particular levels of transparency. The effective creation of transparency is a balancing act between data utility and privacy. To provide a concrete example, let us take the case where one bank is wanting to attain credit scoring models from a third-party vs. another bank which wants to assure its customers that decisions are being made in a fair, regulatory-compliant manner from a third-party. In the first case, data synthesis should be done which maximizes utility over privacy. In this case, the bank would want to be sure that each vendor has as much information as necessary to train the best model possible as the bank will ultimately use that model in production. In the second case, privacy would be prioritized over utility: the bank would like to retain a competitive advantage against a possible bad actors which could seize upon that information to improve their existing models.

% A multi-party mechanism to inform data-owners to guide the data-collection process.
Fourth, many experts can agree that human-mediated processes used to collect data can be inherently biased. Having the ability to share data in a secure manner would be good insurance against biased data collection efforts. Maintaining a diverse set of third-party evaluators becomes critical.
Thus, we wish to make inroads in addressing the inherent biasness and skewness in data. 
This is also linked with the lack of an approach  for  other  researchers  to  reproduce  studies and cross-examine a particular dataset.
%As of late, there is an epidemic of studies not being able to be replicated. 
Having a controlled mechanism for reproducing data is a fundamental element of a model/data sharing solution that combats the issue faced by many studies of reproducability.
%

%

% A compact manner to represent storage-intensive datasets.
%As datasets scale and the internet moves towards a distributed architecture, the replacement for centralized storage may very well be data models. Instead of storing multiple copies of massive databases on edge devices, software engineers my find that the storage of model parameters---model parameters used to regenerate that data---much more economical.

Lastly, we wish to empower users to ensure that their data is forgotten. The implementation of the {\it The Right to Be Forgotten} legislation ensures individuals (or groups of individuals) who choose to no longer be apart of a platform, that their data can be deleted completely. However, user-data removal can impact performance especially in the case of algorithms like collaborative filtering which are used in recommendation engines. We look to mitigate this by using the data-synthesis models instead of the actual data %@kommy -- this is the correct reference. {\color{red}{
(Section \ref{sec:data}). We also track and manage both the data-synthesis models and actual data models (see Section \ref{sec.tme}). %@kommy -- this is the correct reference@Regi/Andrew/Isaac: Is this a correct cross-reference?)}}
%TME, powered by a blockchain network.

%%%%%%%%%%%%%%%%%%%%%%%
\section{Trusted Model Executor}
\label{sec.tme}
%Trusted Model Executor (TME) is a system that evaluates data quality as well as trained model performance upon upload and throughout the model's lifecycle.  
%The Trusted Model Executor (TME) is designed to execute and evaluate the performance of models without the need to disclose the proprietary structural design of the models. The model lifecycle managed by a workflow is recorded on the blockchain. Every action on a model is recorded as an blockchain event. Before a model is executed, the model file is verified using a hash value stored in the blockchain.
%{\color{red}{TODO:}
%\begin{itemize}
 %   \item {\sout{brief description about TME (using the SME network as a motivating FSS network);}} 
 %   \item {\sout{operating model of TME}} 
  %  \item model life cycle using TME.
%\end{itemize}}

\begin{figure}[ht!]
	\centering
	\includegraphics[width=1\ScaleIfNeeded]{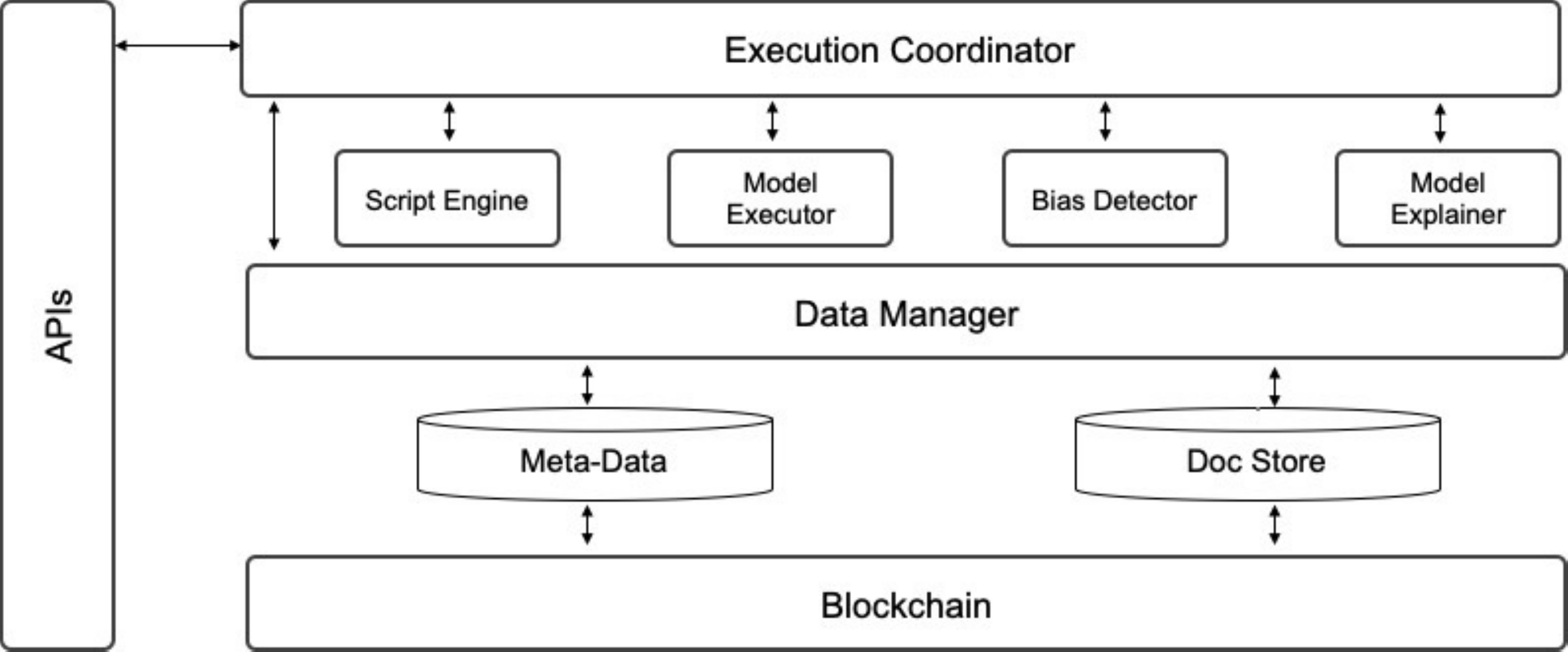}
	\caption{TME Overview.}
	\label{fig:system}
\end{figure}

To address (some of) the above challenges, 
%mentioned in the previous section, 
we developed a mechanism and infrastructure 
to evaluate data quality and trained machine 
learning models throughout their lifecycle.  
These capabilities are presented in what we 
call the Trusted Model Executor (TME) %. 
as shown in Fig. \ref{fig:system}. 
The TME has been 
integrated and tested on a blockchain-based platform for small-to-medium businesses (SMEs). Networked 
digital trust is established among stakeholders along the SME value chain. \cite{%Twiga,
KinaiMOD17}. Each participant involved in the network can interact, view, and/or act on data, models and information pertaining to order contract transactions and decision-making.  

%\begin{comment}
%\end{comment}

\begin{figure*}[ht!]
	\centering
	\includegraphics[width=1\ScaleIfNeeded]{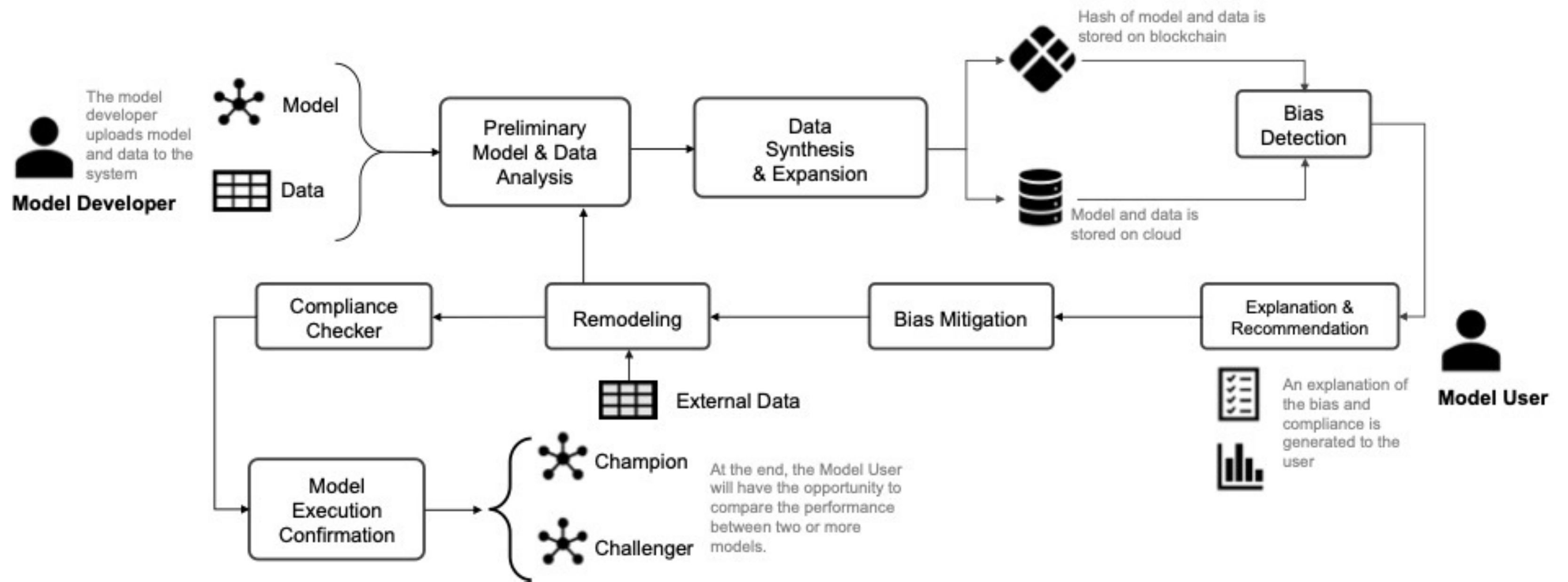}
	\caption{Illustrating the lifecycle of a model.
}
	\label{fig:lifecycle}
\end{figure*}

The TME is designed to execute and evaluate the performance of models (e.g., credit scoring models) without the need to disclose the proprietary structural design of the models. The model lifecycle (Fig. \ref{fig:lifecycle}), is managed by a blockchain controlled workflow, and recorded on the blockchain. Every action on a model is recorded as a blockchain event or sequence of events for transparency and immutability. Before a model is executed, the model file is verified by executing smart contracts.

%\begin{figure}[h!]
%	\centering
%		\includegraphics[width=1\columnwidth]{images/model-life-cycle.pdf}
%%	\label{fig:lifecycle}
%\end{figure}

%\subsection{Model Execution}

Models built using frameworks and languages like Python, PMML, and PKL are accepted and executed by the TME. While pre-processing scripts are used to perform any pre-processing tasks that need to be done on input data before being ran on the model (e.g., removing unnecessary or blank fields), the post-processing scripts performs any extra tasks to the output of the model (e.g., formatting of the model output).  The model execution then provides model explainability of how the various features in the model contributed to the model output. The model execution also supports action triggers, initiated when certain model results are achieved given a set of input parameters. For example, a model that scores businesses can have a trigger that creates a notification whenever small businesses (identified by sales volume) get disproportionately lower credit scores as compared to their larger counterparts.

%\subsection{Model Evaluation} 
The TME supports bias detection and mitigation for both training data and pretrained models by utilizing the underlying capabilities of IBM's AIF360 library \cite{aif360-2018}. For data attributes found with bias based a set of metrics, several mitigation measures can be performed at the user's discretion. Additinally, using a series of model approximation techniques, the TME is able to generate non-expert explanations as to possible causes of the bias in both the dataset and model.

Model/data uploaded to the TME is pre-processed and ingested by the data synthesis module. This module enables users to generate a similar dataset while maintaining the privacy of the original data. The module can also be used to expand the data in cases where data is limited. This module enables sharing of data between users on the TME platform while preserving the privacy of the original data. This will be the main focus for this work.

%The TME supports the evaluation of models leveraging the AIF360 library \cite{aif360-oct-2018}. Training and testing data used to develop the model is uploaded to the TME and then bias detection and mitigation is run on the model. To support this feature of uploading data to the TME, the data synthesis module enables users to upload their data and have a similar data set generated that can be stored by the TME. This can also be used to enable sharing of data between users in the TME.

\section{Data Synthesis and Expansion}
\label{sec.studies}
\begin{comment}
{\color{red}{(@Regi: which box(es) are we ticking from Fig. \ref{fig:lifecycle}?)}}

\textcolor{blue}{RB: The \textit{Data Synthesis \& Expansion} box should be shade. For the purpose of referencing in text, should we enumerate the boxes as well?}
\end{comment}

In this section, we describe the reference datasets used, the generation of data and our experimental studies and analyzing bias-detection utility in synthesized datasets. % and hence no real/private data is necessary {\color{red}(This needs come out from the conclusion of our experiment.)}. 

\subsection{Reference Datasets}\label{sec:data}
We used two tabular datasets for our experimental studies. The purpose of these studies is to demonstrate that the synthetic data generated is able to retain all of the high-level relational information of the real datasets without it being a one-to-one mapping of records from the synthetic data into the real. With that, we would be able to generate synthetic/synthesized data that has similar properties to the real data without any of the privacy risks.

The two datasets are:  
 %\subsubsection{US Adult Census Dataset}

\begin{itemize}
    \item {\it US Adult Census Dataset \cite{USAdult}}. %\footnote{The dataset can be downloaded from \url{https://archive.ics.uci.edu/ml/datasets/adult}}}
    This dataset consists of several personally-identifiable attributes with labels of yearly income. For the preliminary experiments we used a subset of categorical and ordinal variables, such as, work class, education, marital status, occupation, relationship, ethnicity, gender and the target class (income exceeds a threshold). The training dataset contained 32561 records, 7 attributes (listed above) and 1 binary target label.

%\subsubsection{Bank of Portugal Dataset}
\item {\it Bank of Portugal Dataset \cite{Moro2014}}. This consists of 41,188 labeled records with 20 labeled attributes which contain data on whether a customer will accept the term deposits of this particular bank. The preliminary experiments using this dataset used a subset of categorical and ordinal variables, such as, job, marital, education, default, housing, loan, contact, month, last contact day, outcome of previous marketing campaign. The training dataset comprised 37069 records, 10 attributes (listed above) and 1 binary target label.

\end{itemize}

We split these datasets into a training, validation and testing set (70\% training, 10\% validation and 20\% testing). This was selected using a random permutation cross-validation iterator. 
%\subsubsection{Uganda and Kenyan Credit Dataset}
%This dataset consists of 100,000 labeled records with 

\begin{comment}

A diverse set of tabular datasets were chosen—diverse in size as well as number and type of attributes. This selection was done intentionally in order to run experiments that can lead to generalizable results in terms of scoring data skewness and attribute bias with universal metrics. As evidenced by the selection of the targeted attribute, scoring data skew and bias not only have application in ensuring mandated protected attributes (e.g., race, gender) are minimized for bias, but other attributes like regional bias (Uganda (UG) vs. Kenya (KE)) and bias in financial instrument campaign methods (cellular vs. telephone) are properly accounted for so as to prevent financiers from missing out on opportunities in “minority markets” or expose themselves needlessly to avoidable risks. 
\end{comment}

%\subsection{Data Preprocessing}

\subsection{Data Generation}

There are several generative models that can be used to synthesize simulated tabular data that preserves statistical similarity to the original dataset yet prevents information leakage.
Examples of these models are Variational Auto-Encoders (VAE) \cite{kingma2013auto}, Generative Adversarial Networks, and %\cite{goodfellow2014generative}
Dimension-Reduction methods and Kernel Density Estimation. They typically generate new samples that follow the same probabilistic distribution of a given training dataset with a reduced feature vector. We describe the approach used in this work to synthesize the five variations of the two experimental datasets described in Section~\ref{sec:data}.
We used the VAE to develop our data generation method for the purpose of this work. VAE provides a probabilistic mechanism to describe an observation in a latent space. Rather than building an encoder which outputs a single value, our encoder describes a probability distribution for each latent attribute. The entire network is trained as a whole, with two hidden layers for the encoder, two hidden layers for the decoder and the bottle neck layers size is $C \times D$, where $C$ is the number of classes and $D$ the number of categorical distributions. The loss function is the addition of cross-entropy between the output and the input known as the reconstruction loss and the Kullback–Leibler divergence. We trained a standard categorical to generate the samples for both datasets and all the variations. In our case, we use Adaptive Moment Estimation (ADAM) \cite{kingma2014adam} as an optimization method, which computes
adaptive learning rates for each parameter. The input shape of the vectors varies depending on the dataset, and all variables were encoded using one-hot encoding procedure.
 %which penalizes the network for creating outputs different from the input and Kullback–Leibler  divergence which measures the difference between two probability distributions.

%\subsubsection{Training Setup}
%\textcolor{red}{not sure if this section is needed, Reggi to decide}
%Following \cite{jang2016categorical} we trained a standard Categorical VAE to generate the samples for both datasets and all the vairations \textcolor{red}{Reggi can explain the filters applied to the original data}. 
%In our case we use Adaptive Moment Estimation (ADAM) \cite{kingma2014adam} as optimization method, which computes
%adaptive learning rates for each parameter with a $LR=0.001$, $\beta _1 = 0.9$, $\beta _2 = 0.999$. The loss function is the reconstruction loss added to the KL Divergence. The input shape of the vectors varies depending the dataset, all variables were encoded using One hot encoding procedure.

\begin{figure}[!tbp]
    \begin{tabularx}{0.7\columnwidth}{ll}
        \vspace{0cm}
        \hspace{0cm}
         \includegraphics[width=0.4\ScaleIfNeeded]{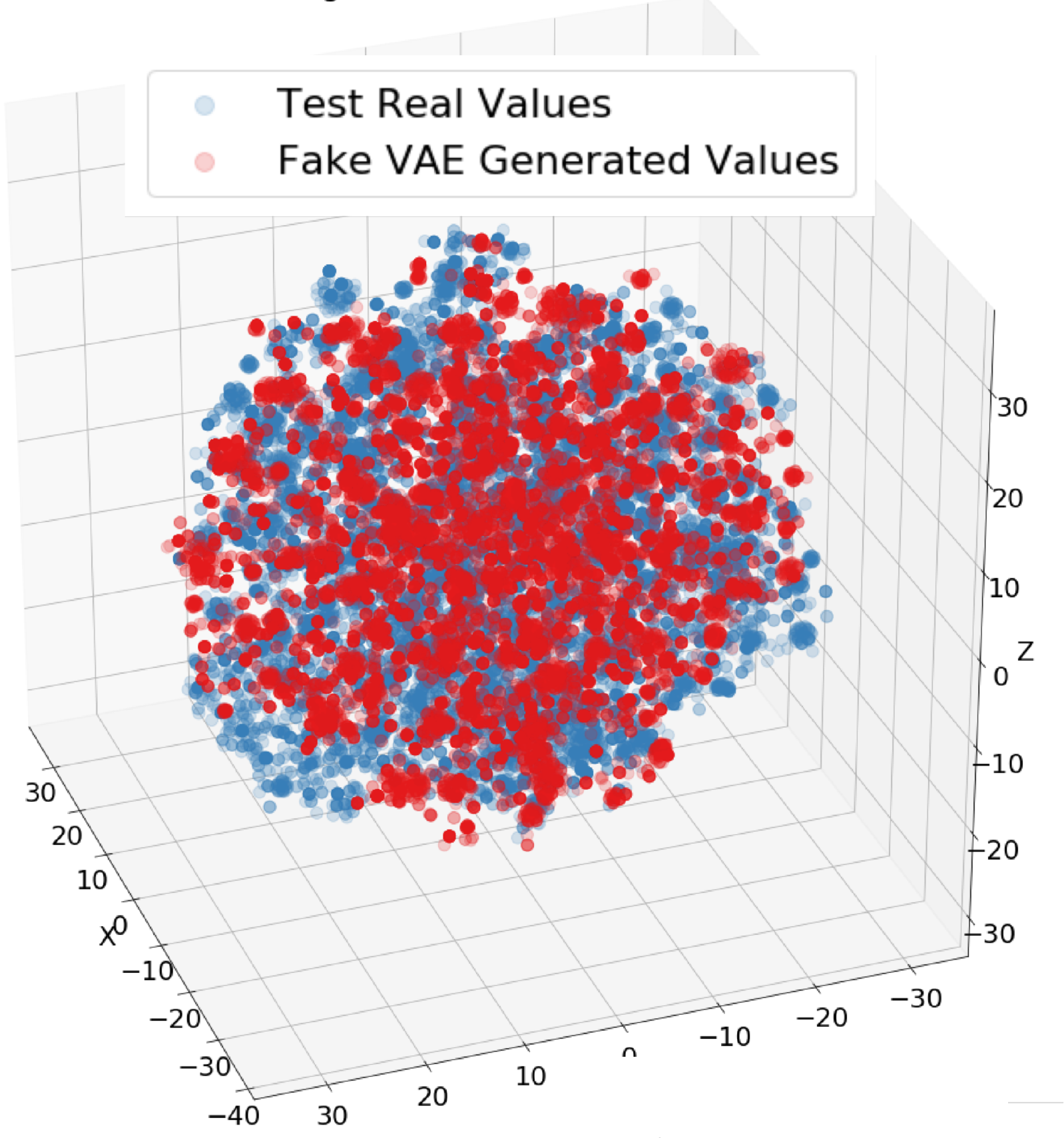}
         &
         \vspace{0cm}
         \hspace{0cm}
         \includegraphics[width=0.4\ScaleIfNeeded]{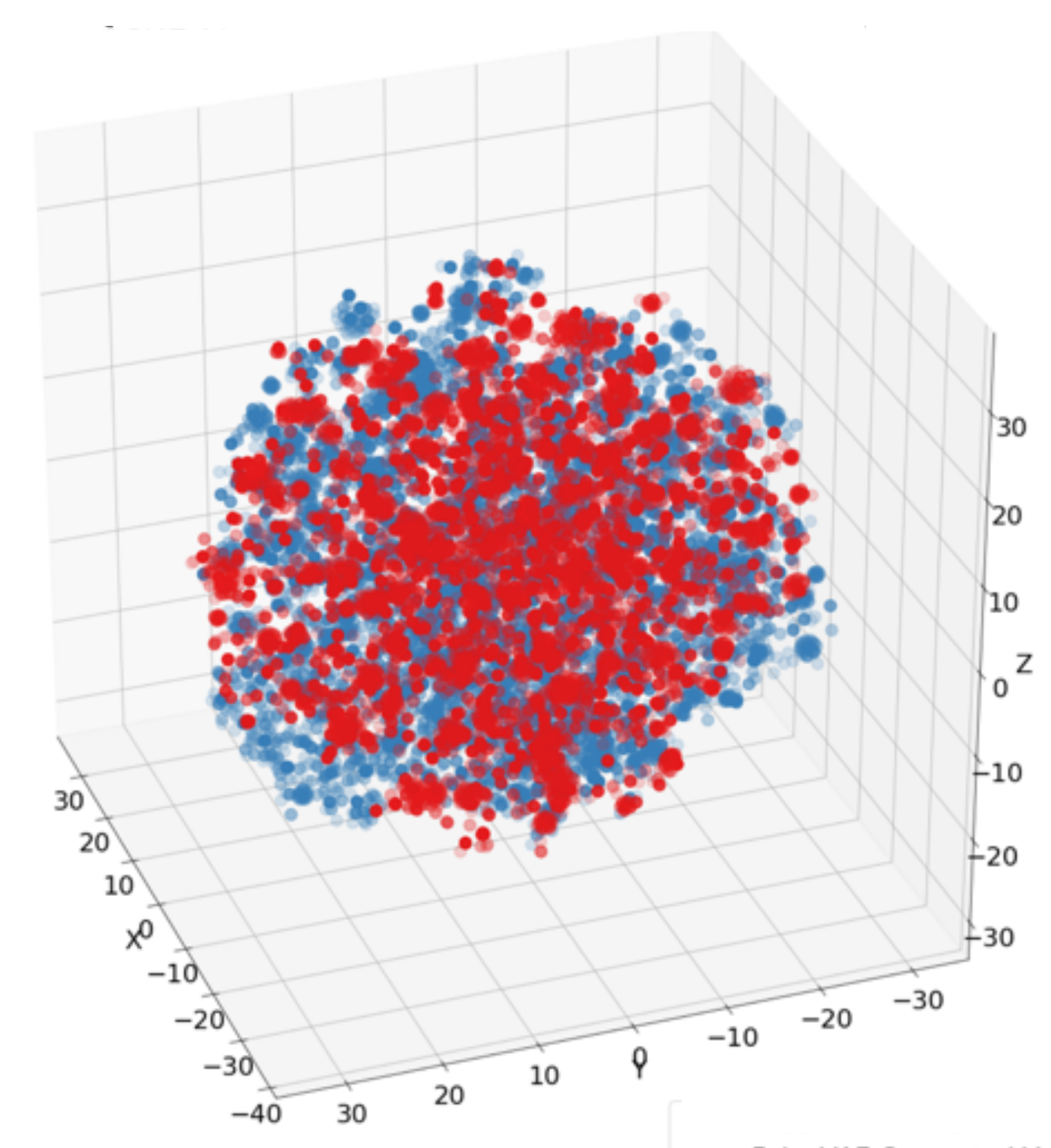}
    \end{tabularx}

  \caption{The feature representation of the raw data distribution using t-SNE is shown with different perspectives. Colors represent the source of the data (real or generated), where red represents generated data and blue real data.}
  \label{fig:t_sne}
\end{figure}

%\subsubsection{Visualization of Generated Data}
Once the data is generated, it is important to understand the representation of the data. We therefore displayed the feature representation of the real and simulated data distribution using t-distributed Stochastic Neighbor Embedding (t-SNE) \cite{VanDerMaaten2008}. t-SNE is an enhanced method for representing high dimensional data by giving each data point a location in a three dimensional map. This can be seen in Fig.~\ref{fig:t_sne}.
\begin{comment}

\begin{figure}[!tbp]
    \begin{subfigure}[b]{0.5\columnwidth}
      \includegraphics[width=\textwidth]{images/1.pdf}
      %\caption{Perspective A}
    \end{subfigure}
    \begin{subfigure}[b]{0.5\columnwidth}
  	    \includegraphics[width=\textwidth]{images/4.pdf}
  	    %\caption{Perspective B}
  	 \end{subfigure}

  \caption{The feature representation of the raw data distribution using t-SNE is shown with different perspectives. Colors represent the source of the data (real or generated), where red represents generated data and blue real data.}
  \label{fig:t_sne}
\end{figure}
\end{comment}

\begin{comment}

\begin{figure}[!h]
 
\begin{subfigure}{0.5\columnwidth}
\includegraphics[width=\ScaleIfNeeded]{images/1.pdf} 
\caption{Caption1}
\label{fig:subim1}
\end{subfigure}
\begin{subfigure}{0.5\columnwidth}
\includegraphics[width=\ScaleIfNeeded]{images/4.pdf}
\caption{Caption 2}
\label{fig:subim2}
\end{subfigure}
 
\caption{Caption for this figure with two images}
\label{fig:image2}
\end{figure}
\end{comment}

\begin{comment}

\begin{figure}[!tbp]
  \centering
  \subfloat[tSNE_A]
  {\includegraphics[width=0.5\ScaleIfNeeded]{images/1.pdf}
  }
  \hfill
  \subfloat[Flower two.]
  {\includegraphics[width=0.5\columnwidth]{images/4.pdf}
  }
  \label{fig:t_sne}
\end{figure}
\end{comment}

\subsection{Data synthesis and experimental bias evaluation}

% Attributes == Features

Our studies seek to experimentally identify and characterize the set of bias metrics that should be tracked for the synthetic data. For this work, we are performing a comparative analysis between the real and synthetic datasets using the following metrics
\begin{itemize}
  \item Statistical Parity Difference (Stat. Diff.)
  \item Disparate Impact (Disp. Imp.)
  \item K-Nearest Neighbors Consistency (Consistency) \cite{dwork2012fairness}
  \item Number of Positive Examples (Num. Neg.)
  \item Number of Negative Examples (Num. Neg.)
  \item Base Rate
  
\end{itemize}

For this work, the analysis focused on between group fairness metrics as determined by statistical parity difference. and Disparate impact as well as the individual fairness metric captured by data consistency. 

Statistical parity difference is defined as:

\begin{align}
%\Pr \left( Y = 1 | D = \text{unpriv} \right) - \Pr (Y = 1 | D = \text{priv})
\Pr \left( y = 1 | b \in dom(S) \right) - \Pr (y = 1 | w \in dom(S) )
\label{statdiff}
\end{align}

Disparate impact is defined as:

\begin{align}
%\frac{\Pr (Y = 1 | D = \text{unprivileged})} {\Pr (Y = 1 | D = \text{privileged})}
\frac{\Pr (y=1 | b \in dom(S))} {\Pr (y=1 | w \in dom(S))} 
\label{dispimp}
\end{align}

Consistency is defined as:
\begin{align}
1 
- \frac{1}{N\text{k}}
\sum_{i=1}^N 
\left | y_i - 
\sum_{j\in\mathcal{N}_{\text{kNN}}(x_i)}y_j   \right |
\label{consistency}    
\end{align}

For Equations (\ref{statdiff}), (\ref{dispimp}) and \ref{consistency}, we assume that the labled datasets are defined by $(X, Y)$, where $X = \{ x_i, ..., x_N \}$ is the set of attributes and $Y = \{y_i, ..., y_N \}$ the labels. Generally, the domain of $X$, $dom(X)$, can take on a variety of data types. As stated, for our analysis, $dom(X)$ is restricted to categorical (nominal) and ordinal values with low cardinality. Moreover, the domain of $Y$ is restricted to binary label classes: $y \in dom(Y) = \{ 1, 0\}$. For bias, a single attribute in $X$ is designated as the $S$ \textit{sensitive attribute}: $S \in  X$. For our analysis, $S$ also takes on binary values, $dom(S) = \{b,w\}$, where $b$ is designated as the unprivileged class and $w$ the privileged. For our experiements $S$ was set to gender and contact method (contact) in the US Adult Census and Bank of Portugal datasets, respectively, with $b$ and $w$ set to $dom(\text{gender}) = (\text{female},\text{male})$ and $dom(\text{contact}) = (\text{land line telephone}, \text{cellular})$.

For Equation (\ref{consistency}),  $\mathcal{N}_{kNN}$ is the k-Nearest Neighbor function used to identify $k$-number ($k=5$, in our case) of instance around $x_i$ in attribute space. Ideally, those five neighbors should have the same label as $x_i$. Any discrepancies will reduce a perfect consistency score of one.

%% ================
\begin{comment}
Base Rate:
\begin{align}
\Pr(Y=1) = P/(P+N)    
\end{align}

Num. Neg.:
\begin{align}
    NumPos = \sum_{i=1}^n \mathbbm{1}[y_i = 0]
\end{align}

Num. Pos.:
\begin{align}
    NumPos = \sum_{i=1}^n \mathbbm{1}[y_i = 0]
\end{align}
\end{comment}
%% ================

Number of positive instances, $ \sum_{i=1}^n \mathbbm{1}[y_i = 1] $, number of negative instances, ($ \sum_{i=1}^n \mathbbm{1}[y_i = 0] $) and base rate, ($\text{Num. Pos.}/N$) represent the unconditioned class probabilities of the labels. Naturally as each of the datasets are intentionally skewed, those three metrics are expected to change accordingly.

\begin{figure}[h!]
    \centering
    \includegraphics[width=0.9\ScaleIfNeeded]{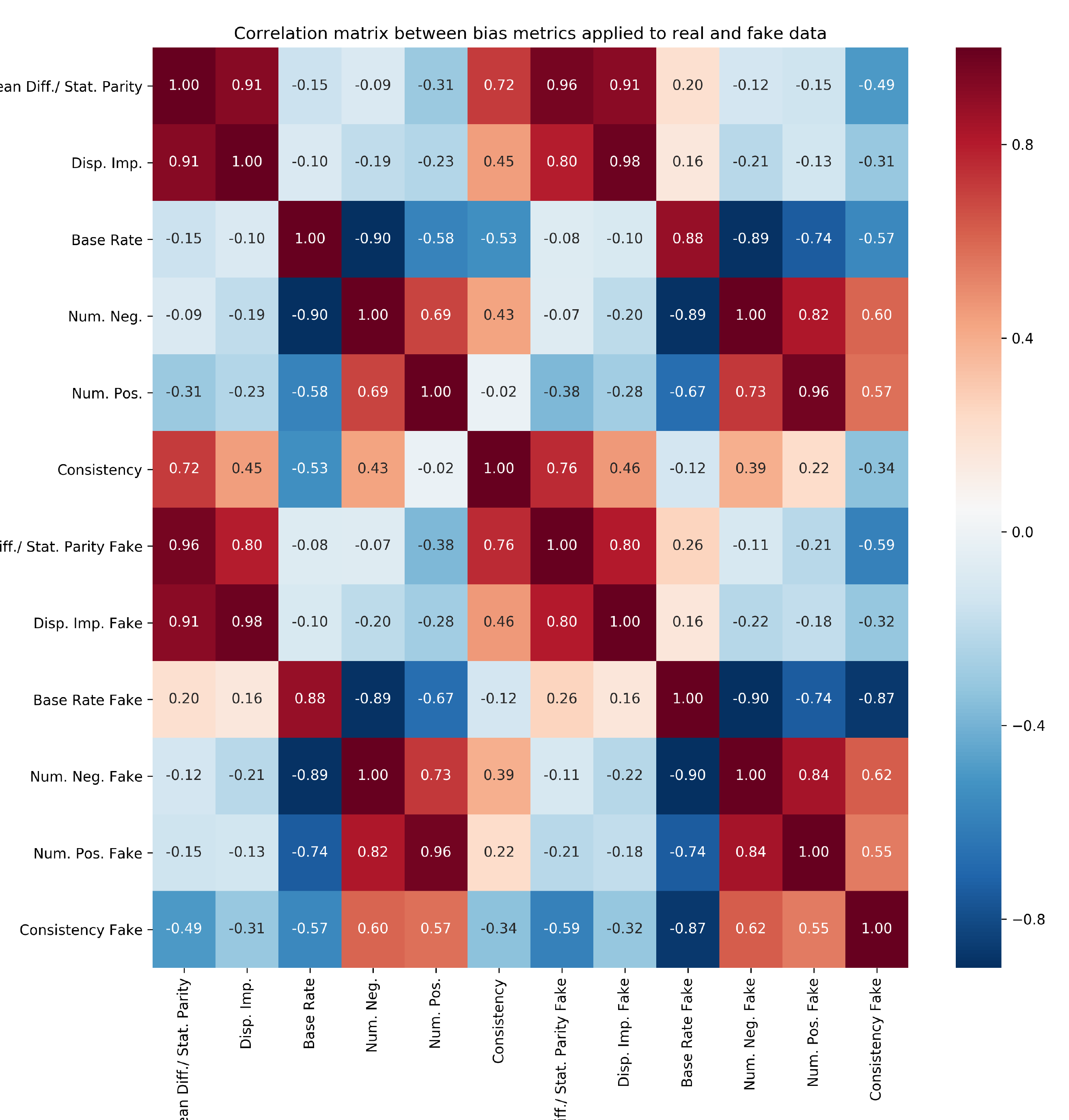}
    \caption{Correlation matrix comparing the features and bias (Statistical Parity Difference and Disparate Impact) metric scores of the real and generated dataset variations. There is high correlation in four out of the five bias metrics. $Stat. Parity Diff = 0.96$,  $Disp. Imp. = 0.98$, $Base Rate = 0.88$,
$Num. Neg = 1.00$,  $Num. Pos. = 0.96$ and $Consistency = -0.34$.}
    \label{fig:corr}
\end{figure}

Figure ~\ref{fig:corr} is a correlation map for both the real and synthetic (synthetic) datasets. Examining top off-axis correlations between the synthetic and real datasets, the top off-axis correlations are shown to be between the number of all negative instances (Num. Neg.), $1.00$, the disparte impact (Disp. Imp.), $0.98$ and statistical parity difference (Stat. Parity Diff.). Conversely, the consistency metric shows a weak relationship between the real and synthetic datasets.

\subsection{Analysis and Results}

Upon characterizing and scoring of each of the 20 datasets, several initial trends emerged. Figure \ref{fig:corr} highlights these trends in a correlation map among the five selected metrics. Worth noting is the real to synthesized dataset correlation for the Disp. Imp. and Stat. Parity Diff. metrics. The high correlation suggests that our method for data reproduction should be able to preserve group bias while effectively breaking the one-to-one connection between the original and synthesized datasets. It should be noted that while both datasets tracked the monotonic trends of increasing and decreasing DI, the synthesized data in the Bank Portugal dataset experiments tracked the scale changes much more closely to the real one when compared against the Adult USA experiments. More fine grain dataset sampling and more dataset types are required to fully understand the underline behavior of VAE-generated data with respect to group bias tracking.  

\section{Discussion and Summary}
In this work, we presented our implementation of a trusted model-lifecycle management platform, highlighting the \textit{Data Synthesis and Expansion} module. Specifically, the focus was on how to securely distribute datasets (containing sensitive information) to third-party evaluators by using Variational Auto-Encoder (VAE) technology. The goal was to generate synthetic data from the latent representation of the original data in order to preserve privacy while retaining the utility of that original data. In our case, the utility of bias detection in the synthetic dataset was measured using the bias in the original dataset as the ground truth. Several bias metrics including group and individual bias were examined as two financial datasets were artificially skewed by a subsampling process. Experimentally, our results lead us to believe that using the VAE for data reproduction can effectively retain some of the high-level statistical information from the original dataset. However, individual bias may not be retained during the data reproduction process. 

More datasets and experimental evaluations are required in order to uncover the relationship that may exist between real and VAE-generated tabular data.

\bibliographystyle{IEEEtran}
\bibliography{main}
\end{document}